\documentclass[pra,superscriptaddress,showpacs, twocolumn,10pt]{revtex4}
\usepackage{amssymb}
\usepackage{graphicx}

\def\>{\rangle}

\begin{document}

\newtheorem{corollary}{Corollary}
\newtheorem{definition}{Definition}
\newtheorem{example}{Example}
\newtheorem{lemma}{Lemma}
\newtheorem{proposition}{Proposition}
\newtheorem{theorem}{Theorem}
\newtheorem{fact}{Fact}
\newtheorem{property}{Property}
\renewcommand{\choose}[2]{{{#1}\atopwithdelims(){#2}}}

\title{Entanglement-assisted transformation is asymptotically equivalent to multiple-copy transformation}
\author{Runyao Duan}
\email{dry02@mails.tsinghua.edu.cn}
\author{Yuan Feng}
\email{feng-y@tsinghua.edu.cn}
\author{Mingsheng Ying}
\email{yingmsh@tsinghua.edu.cn}

\affiliation{State Key Laboratory of Intelligent Technology and
Systems, Department of Computer Science and Technology, Tsinghua
University, Beijing, China, 100084}

\date{\today}

\begin{abstract}
We show that  two ways of manipulation of quantum entanglement,
namely, entanglement-assisted local transformation [D. Jonathan
and M. B. Plenio, Phys. Rev. Lett. {\bf 83}, 3566 (1999)] and
multiple-copy transformation [S. Bandyopadhyay, V. Roychowdhury,
and U. Sen, Phys. Rev. A {\bf 65}, 052315 (2002)], are equivalent
in the sense that they can asymptotically  simulate each other's
ability to implement a desired transformation from a given source
state to another given target state with the same optimal success
probability. As a consequence, this yields a feasible method to
evaluate the optimal conversion probability of an
entanglement-assisted transformation.
\end{abstract}

\pacs{03.67.Hk, 03.67.Mn}

\maketitle

As a valuable resource in quantum information processing, quantum
entanglement has been widely used in quantum cryptography
\cite{BB84}, quantum superdense coding  \cite{BS92}, and quantum
teleportation \cite{BBC+93}. Consequently, it remains the subject
of interest at present after years of investigations.

Unlike common resources, it was discovered  by Jonathan and Plenio
\cite{JP99} that quantum entanglement is truly a strange one:
sometimes it can help quantum information processing without being
consumed at all. This effect can be understood in the situation of
entanglement transformation. Suppose that two spatially separated
parties, say, Alice and Bob, share a finite dimensional entangled
pure state $|\psi_1\rangle$, and they want to convert
$|\psi_1\rangle$ into another state $|\psi_2\rangle$, by using
local quantum operations and classical communication (LOCC) only
\cite{BBPS96, NI99, Vidal99,JP99, JW00, VC02, BPRST00,DK01,
SRS02,FDY04, FDY04a, DFLY05, DFY05a}. For certain $|\psi_1\rangle$
and $|\psi_2\rangle$, they can accomplish their goal with
certainty by constructing a local protocol \cite{NI99}. While in
general, only a maximal conversion probability less than one can
be achieved \cite{Vidal99}. In the latter case, Jonathan and
Plenio demonstrated by examples that sometimes Alice and Bob may
borrow another entangled state $|\phi\rangle$, known as a
catalyst, to realize  the transformation $|\psi_1\rangle
\rightarrow |\psi_2\rangle$ with probability one. The
transformation can be represented as
$|\psi_1\rangle\otimes|\phi\rangle \rightarrow
|\psi_2\rangle\otimes|\phi\rangle$, in which it is obvious that
the catalyst $|\phi\rangle$ is not consumed during the process.
Such a transformation that uses intermediate entanglement without
consuming it is called `entanglement-assisted local
transformation' in Ref. \cite{JP99}, abbreviated to ELOCC. The
mathematical structure of ELOCC has been studied thoroughly in
Refs. \cite{DK01,FDY04,FDY04a}. It has also been shown that such
an entanglement catalysis phenomenon exists in the manipulation of
mixed states \cite{JW00}, and in the implementation of non-local
quantum operations \cite{VC02}.

Another interesting way of manipulating  quantum entanglement was
proposed by Bandyopadhyay $et\ al$ \cite{SRS02}. Specifically,
they found that sometimes multiple copies of source state may be
transformed into the same number of target state although the
transformation cannot happen for a single copy. That is, for some
states $|\psi_1\rangle$ and $|\psi_2\rangle$, although Alice and
Bob cannot transform $|\psi_1\rangle$ into $|\psi_2\rangle$ with
certainty by LOCC, there may exist $m>1$ such that they can
realize the transformation $|\psi_1\rangle^{\otimes m}\rightarrow
|\psi_2\rangle^{\otimes m}$ with certainty. This kind of
transformation that uses multiple copies of source state and then
transforms all of them together into the same number of target
state is called `non-asymptotic bipartite pure-state entanglement
transformation' in \cite{SRS02}. More intuitively, it can also be
called `multiple-copy entanglement transformation', or MLOCC for
short \cite{DFLY05a}. The mathematical structure of MLOCC was
carefully examined in Ref. \cite{DFLY05}.

 At  first glance,  entanglement-assisted transformation and multiple-copy
entanglement transformation are two completely different
extensions of ordinary LOCC. To achieve a specific transformation,
the former needs to borrow extra entanglement as resource but is
promised  not to consume it during the transformation, while the
latter realizes a similar purpose by accumulating a sufficiently
large number of copies of source state and then transforms all
these copies together into the same number of target state.

A surprising fact is that these two kind of manipulations of
entanglement are closely related to each other. In Ref.
\cite{DFLY05} it was demonstrated that if a bipartite entangled
state $|\psi_1\rangle$ can be transformed into another bipartite
entangled state $|\psi_2\rangle$ with certainty by using MLOCC,
then $|\psi_1\rangle$ can also be deterministically transformed
into $|\psi_2\rangle$  by using a catalyst. In other words, in the
deterministic scenario ELOCC is at least as powerful as MLOCC.
However, this result is of limited interest because in general for
any two given states $|\psi_1\rangle$ and $|\psi_2\rangle$ the
transformation of $|\psi_1\rangle$ to $|\psi_2\rangle$ cannot be
achieved with certainty by using ELOCC \cite{JP99}. It is a very
interesting problem to further explore the precise relation
between ELOCC and MLOCC.

An interesting equivalence of ELOCC and MLOCC was observed by the
authors previously  in Ref. \cite{DFY05a}, where we only concerned
that whether ELOCC or MLOCC transformations have advantages over
pure LOCC ones in producing a given target. Specifically, for an
$n\times n$ target state $|\psi_2\rangle$ and a positive integer
$k$, if there exists another $n\times n$ state $|\psi_1\rangle$
such that the transformation of $|\psi_1\rangle$ to
$|\psi_2\rangle$ can be achieved with certainty by using a
$k\times k$-dimensional catalyst while $|\psi_1\rangle$ cannot be
transformed into $|\psi_2\rangle$ by LOCC, then we say $k$-ELOCC
is useful in producing $|\psi_2\rangle$. The concept that
$k$-MLOCC is useful in producing a given target state can be
defined in a similar way. Then we proved that $k$-ELOCC  is useful
in producing $|\psi_2\rangle$ if and only if $k$-MLOCC is useful
in producing the same target. An explicit necessary and sufficient
condition for both of them in terms of Schmidt coefficients of
$|\psi_2\rangle$ was also obtained. This equivalence was then
generalized to probabilistic transformations. See Theorems 2 and 4
in Ref. \cite{DFY05a} for details. Obviously, this kind of
equivalence is weak in the sense that only the target state is
involved, and the source state is irrelevant.

In this brief report, we consider probabilistic transformations
instead of deterministic ones and we obtain a very strong
equivalence between ELOCC and MLOCC. We find that ELOCC and MLOCC
are indeed equivalent in the sense that they can simulate each
other's ability to implement any given transformation with the
same optimal success probability. More precisely, let
$|\psi_1\rangle$ and $|\psi_2\rangle$ be the source state and the
target state  of a transformation, respectively. We further assume
that $P_E(|\psi_1\rightarrow |\psi_2\rangle)$ (see Eq. (\ref{pe})
bellow) is the optimal success probability that can be achieved by
using some catalyst. Similarly, $P_M(\psi_1\rightarrow
|\psi_2\rangle)$ (see Eq. (\ref{pm}) bellow) is the optimal
average probability that can be achieved by transforming multiple
copies of $|\psi_1\rangle$ into the same number of copies of
$|\psi_2\rangle$. Then we prove that
$P_E(|\psi_1\rightarrow|\psi_2\rangle)$ is exactly the same as
$P_M(|\psi_1\rightarrow |\psi_2\rangle)$ for any choices of
$|\psi_1\rangle$ and $|\psi_2\rangle$. It is clear that this
equivalence is very different from the one obtained in Ref.
\cite{DFY05a}. In fact it is much more elaborate since it
completely characterizes the equivalence of ELOCC and MLOCC in the
probabilistic scenario.

This equivalence of ELOCC and MLOCC transformations is interesting
in many ways, both theoretically and practically. In principle, it
uncovers an essential connection between entanglement catalysis
and multiple-copy entanglement transformation, and declares that
they have almost the same effect. In practice, it provides a more
feasible way to evaluate the optimal conversion probability of an
ELOCC transformation by calculating the optimal conversion
probability of the corresponding MLOCC one. The proof presented in
this brief report also reveals that such an equivalence is deeply
related to a well-known fact: a maximally entangled state cannot
serve as a catalyst, which puts a fundamental constraint on the
power of entanglement-assisted transformation.

Let us begin with a concrete example to examine the relationship
between entanglement-assisted transformation and multiple-copy
entanglement transformation. The primary tool required for this is
Vidal's formula \cite{Vidal99}. Let
$|\psi_1\rangle=\sum_{i=1}^n\sqrt{\alpha_i}|i_A\rangle|i_B\rangle$
and
$|\psi_2\rangle=\sum_{i=1}^n\sqrt{\beta_i}|i_A\rangle|i_B\rangle$
be pure bipartite entangled states with ordered Schmidt
coefficients $\alpha_1\geq \alpha_2\geq \cdots \geq \alpha_n>0$
and $\beta_1\geq \beta_2\geq \cdots\geq \beta_n\geq 0$,
respectively. Then the maximal conversion probability of
transforming $|\psi_1\rangle$ into $|\psi_2\rangle$ by LOCC is
given by \cite{Vidal99}
\begin{equation}\label{vidalformula}
p_{max}(|\psi_1\rangle\rightarrow |\psi_2\rangle)={\rm
min}\{\frac{E_l{(|\psi_1\rangle)}}{E_l(|\psi_2\rangle)}:1\leq
l\leq n\},
\end{equation}
where $E_l({|\psi_1\rangle})=\sum_{i=l}^{n}\alpha_i$. In the case
of $p_{max}(|\psi_1\rangle\rightarrow |\psi_2\rangle)=1$, i.e.,
the transformation $|\psi_1\rangle\rightarrow |\psi_2\rangle$ can
be realized with certainty under LOCC,  Vidal's formula reduces to
Nielsen's theorem \cite{NI99}.

Now an example demonstrating the power of entanglement-assisted
transformation  is as follows. Take $
|\psi_1\rangle=\sqrt{0.4}|00\rangle+\sqrt{0.4}|11\rangle+\sqrt{0.1}|22\rangle+\sqrt{0.1}|33\rangle$
and
$|\psi_2\rangle=\sqrt{0.5}|00\rangle+\sqrt{0.25}|11\rangle+\sqrt{0.25}|22\rangle.$
We know that the transformation $|\psi_1\rangle\rightarrow
|\psi_2\rangle$ cannot occur with certainty under LOCC since
$p_{max}(|\psi_1\rangle\rightarrow |\psi_2\rangle)=0.8.$  But if
another entangled state
$|\phi\rangle=\sqrt{0.6}|44\rangle+\sqrt{0.4}|55\rangle$ is
introduced, then the transformation $
|\psi_1\rangle\otimes|\phi\rangle \rightarrow
|\psi_2\rangle\otimes|\phi\rangle $ can be realized with certainty
because $p_{max}(|\psi_1\rangle\otimes |\phi\rangle\rightarrow
|\psi_2\rangle\otimes |\phi\rangle)=1.$ Interestingly, the same
task can be achieved by a multiple-copy entanglement
transformation. It is not difficult to see that the transformation
$|\psi_1\rangle^{\otimes 3}\rightarrow |\psi_2\rangle^{\otimes 3}
$ occurs with certainty by checking
$p_{max}(|\psi_1\rangle^{\otimes 3}\rightarrow
|\psi_2\rangle^{\otimes 3})=1.$ That is, when Alice and Bob
prepare three copies of $|\psi_1\rangle$ instead of just a single
one, they can transform these three copies all together into three
copies of $|\psi_2\rangle$ by LOCC.

In the above example, entanglement-assisted transformation and
multiple-copy entanglement transformation can be simulated by each
other. Indeed, it is not difficult to find more examples in which
the same thing happens. This motivates us to conjecture that these
two kinds of manipulation of entanglement are in fact equivalent
in the sense that they can be simulated by each other in some way.

As will be seen later, every multiple-copy entanglement
transformation can be realized by an entanglement-assisted one.
Unfortunately, the following example shows that sometimes an
entanglement-assisted transformation is  more powerful than a
corresponding multiple-copy entanglement transformation. Let us
take source state and target state as $
|\psi_1\rangle=\frac{1}{\sqrt{1.01}}(\sqrt{0.40}|00\rangle+\sqrt{0.40}|11\rangle
+\sqrt{0.10}|22\rangle +\sqrt{0.1}|33\rangle +
\sqrt{0.01}|44\rangle)$ and
$|\psi_2\rangle=\frac{1}{\sqrt{1.01}}(\sqrt{0.50}|00\rangle+\sqrt{0.25}|11\rangle+\sqrt{0.20}|22\rangle
+\sqrt{0.05}|33\rangle+\sqrt{0.01}|44\rangle),$ respectively. A
simple calculation carries out that $|\phi\rangle$ given in the
above example is a multiple-copy catalyst for the transformation
from $|\psi_1\rangle$ to $|\psi_2\rangle$, since it holds that
$p_{max}(|\psi_1\rangle\otimes |\phi\rangle^{\otimes
11}\rightarrow |\psi_2\rangle\otimes |\phi\rangle^{\otimes
11})=1.$ On the other hand, noticing here that the least Schmidt
coefficients of the source and target are very small and
identical, we can show that $p_{max}(|\psi_1\rangle^{\otimes
m}\rightarrow |\psi_2\rangle^{\otimes m})<1 {\rm\ for\ any\ }m\geq
1$ by a tedious but routine calculation. Therefore, for all $m\geq
1$, it always holds that $p_{max}(|\psi_1\rangle^{\otimes
m}\rightarrow |\psi_2\rangle^{\otimes
m})<p_{max}(|\psi_1\rangle\otimes |\phi\rangle^{\otimes
11}\rightarrow |\psi_2\rangle\otimes |\phi\rangle^{\otimes 11})$.
Such an example shows that an entanglement-assisted transformation
cannot be realized by a multiple-copy entanglement transformation
in a finite manner.

We now turn to observe the relationship between
entanglement-assisted transformation and multiple-copy
entanglement transformation in an asymptotical manner.
Surprisingly, these two kinds of transformation are asymptotically
equivalent although it is not the case when only a finite manner
is allowed. To verify this, we need to introduce several
notations. For each $m\geq 1$, let
\begin{equation}\label{avpm}
p_M^{(m)}(|\psi_1\rangle\rightarrow
|\psi_2\rangle)=[p_{max}(|\psi_1\rangle^{\otimes m}\rightarrow
|\psi_2\rangle^{\otimes m})]^{\frac{1}{m}}.
\end{equation}
Intuitively, $p_M^{(m)}$ is the geometric average value of the
probability of (single-copy) transformation
$|\psi_1\rangle\rightarrow |\psi_2\rangle$ when considering in the
environment of $m$-copy transformation $|\psi_1\rangle^{\otimes
m}\rightarrow |\psi_2\rangle^{\otimes m}$. (Note that it is
reasonable to compare the probability
$p_{max}(|\psi_1\rangle\otimes |\phi\rangle\rightarrow
|\psi_2\rangle\otimes
   |\phi\rangle)$ with $p_M^{(m)}(|\psi_1\rangle\rightarrow
|\psi_2\rangle)$ rather than $p_{max}(|\psi_1\rangle^{\otimes
m}\rightarrow |\psi_2\rangle^{\otimes m})$, because the latter is
the probability that $m$ copies of $|\psi_1\rangle$ are
transformed simultaneously to the same number of $|\psi_2\rangle$
and it is usually the $m$th power of the probability of
single-copy transformation.) Then the optimal conversion
probability of a multiple-copy entanglement transformation is
given by
\begin{equation}\label{pm}
P_M(|\psi_1\rangle\rightarrow |\psi_2\rangle)=\sup_{m}
p_M^{(m)}(|\psi_1\rangle\rightarrow |\psi_2\rangle),
\end{equation}
where $m$ ranges over all positive integers. On the other hand, we
define the optimal conversion probability of an
entanglement-assisted transformation from $|\psi_1\rangle$ to
$|\psi_2\rangle$  by
\begin{equation}\label{pe}
P_E(|\psi_1\rangle\rightarrow |\psi_2\rangle)=\sup_{|\phi\rangle}
p_{max}(|\psi_1\rangle\otimes |\phi\rangle\rightarrow
|\psi_2\rangle\otimes |\phi\rangle),
\end{equation}
where $|\phi\rangle$ ranges over all finite dimensional bipartite
pure states.

Now with the notations introduced above, the asymptotical
equivalence of ELOCC and MLOCC can be exactly stated in the
following:
\begin{theorem}\label{mise}\upshape
For any pure bipartite states $|\psi_1\rangle$ and
$|\psi_2\rangle$, $P_E(|\psi_1\rangle\rightarrow|\psi_2\rangle)=
P_M(|\psi_1\rangle\rightarrow|\psi_2\rangle).$
\end{theorem}

\textrm{Proof.} For simplicity, in this proof we abbreviate
$P_E(|\psi_1\rangle\rightarrow|\psi_2\rangle)$ and
$P_M(|\psi_1\rangle\rightarrow|\psi_2\rangle)$ to $P_E$ and $P_M$,
respectively.

We first prove that $P_E\geq P_M$. In fact, we can prove that any
multiple-copy entanglement transformation can be simulated by  a
suitable entanglement-assisted transformation with a finite
dimensional catalyst state.

For simplicity of notations, for any $m\geq 1$, we denote the
geometric average probability of $m$-copy transformation, namely
$p_M^{(m)}(|\psi_1\rangle\rightarrow |\psi_2\rangle)$ in Eq.
(\ref{avpm}), by $p_m$. We will show that there always exists a
finite dimensional catalyst state $|\phi\rangle$ such that
\begin{equation}\label{catalyst}
p_{max}(|\psi_1\rangle\otimes |\phi\rangle\rightarrow
|\psi_2\rangle\otimes |\phi\rangle)\geq p_m.
\end{equation}
Then $P_E\geq P_M$ follows immediately from Eq. (\ref{catalyst}),
by taking supremums according to $m$ and $|\phi\rangle$,
respectively.

For this purpose, let us construct a catalyst state $|\phi\rangle$
as follows:
\begin{equation}\label{lambdacatalyst}
\lambda(\phi)=\lambda(\psi_1^{\otimes m-1}\oplus
p_m\psi_1^{\otimes m-2}\otimes \psi_2\oplus \cdots \oplus
p_m^{m-1}\psi_2^{\otimes m-1}),
\end{equation}
where we use $\lambda(\phi)$ to denote the ordered Schmidt
coefficient vector of $|\phi\rangle$ and  an unimportant
normalization  factor of $\lambda(\phi)$ is omitted. The intuition
behind such a construction comes from the following well known
algebraic identity:
$$x^m-p^my^m=(x-py)(x^{m-1}+px^{m-2}y+\cdots+p^{m-1}y^{m-1}).$$

Now by using Lemmas 3 and 4 in Ref. \cite{FDY04} repeatedly, we
can easily verify the validity of Eq. (\ref{catalyst}). Thus we
complete the proof of $P_E\geq P_M$.

Conversely, we prove the other part that $P_E\leq P_M$. To this
end, a natural strategy is to show that the role of a catalyst
state $|\phi\rangle$ can always be replaced by a suitable $m$-copy
transformation. Unfortunately, according to the second example
presented above, this idea does not work. To overcome this
difficulty, we try to simulate a catalyst state by using
multiple-copy entanglement transformation in an asymptotical
manner.

Before going into the detailed proof, we describe some basic proof
ideas here.  Two points are crucial in the subsequent proof: the
first one is that we can always simulate any $k\times k$ state
$|\phi\rangle$ by a maximally entangled state
$|\Phi_k\rangle=\frac{1}{\sqrt{k}}\sum_{i=1}^k|i_A\rangle|i_B\rangle$
with a nonzero probability; the second one is that  a maximally
entangled state does not have catalysis effect, which puts a
fundamental constraint on the power of entanglement catalysis.

Let us continue the proof of  $P_E\leq P_M$. Assume that
$|\phi\rangle$  is  a $k\times k$ catalyst with the least Schmidt
coefficient $\gamma_k>0$. To generate such a state, Alice and Bob
first borrow a maximally entangled state $|\Phi_k\rangle$.  Then
they try to obtain a lower bound of the successful conversion
probability of the transformation from $|\psi_1\rangle^{\otimes
m}\otimes |\Phi_k\rangle$ to $|\psi_2\rangle^{\otimes m}\otimes
|\Phi_k\rangle$. A possible protocol implementing this task
consists of the following three steps:

1) Generate catalyst state $|\phi\rangle$. That is, perform the
transformation
$$|\psi_1\rangle^{\otimes m}\otimes |\Phi_k\rangle\rightarrow
|\psi_1\rangle^{\otimes m}\otimes |\phi\rangle.$$  The maximal
successful conversion probability is denoted by $p_1$.  This
transformation  can be realized by  transforming $|\Phi_k\rangle$
directly into $|\phi\rangle$ and keeping $|\psi_1\rangle^{\otimes
m}$ intact. Thus
\begin{equation}\label{p1}
p_1=p_{max}(|\Phi_k\rangle\rightarrow |\phi\rangle)=1.
\end{equation}

2) Catalyze $|\psi_1\rangle^{\otimes m}$ into
$|\psi_2\rangle^{\otimes m}$. That is, perform the transformation
$$|\psi_1\rangle^{\otimes m}\otimes |\phi\rangle\rightarrow
|\psi_2\rangle^{\otimes m}\otimes |\phi\rangle.$$ The maximal
successful conversion probability is denoted by $p_2$. This
transformation can be achieved by repeatedly transforming
$|\psi_1\rangle$ into $|\psi_2\rangle$ $m$ times with
$|\phi\rangle$ serving as a catalyst. Note that the key point here
is that as a catalyst, $|\phi\rangle$ here, is reusable. Thus
\begin{equation}\label{p2}
p_2\geq [p_{max}(|\psi_1\rangle\otimes |\phi\rangle\rightarrow
|\psi_2\rangle\otimes |\phi\rangle)]^m.
\end{equation}

3) Return the maximally entangled state $|\Phi_k\rangle$. That is,
perform the transformation
$$|\psi_2\rangle^{\otimes m}\otimes |\phi\rangle\rightarrow
|\psi_2\rangle^{\otimes m}\otimes |\Phi_k\rangle.$$ The maximal
successful conversion probability is denoted by $p_3$. It is easy
to see that this transformation can be implemented by transforming
$|\phi\rangle$ into $|\Phi_k\rangle$ and keeping
$|\psi_2\rangle^{\otimes m}$ intact. Thus
\begin{equation}\label{p3}
p_3\geq p_{max}(|\phi\rangle\rightarrow |\Phi_k\rangle)=k\gamma_k.
\end{equation}

The above three steps complete a concrete protocol to realize the
transformation from $|\psi_1\rangle^{\otimes m}\otimes
|\Phi_k\rangle$ to $|\psi_2\rangle^{\otimes m}\otimes
|\Phi_k\rangle$. Hence we obtain a lower bound for the maximal
successful conversion  probability of this transformation, that
is,
\begin{equation}\label{lowbound}
p_{max}(|\psi_1\rangle^{\otimes m}\otimes
|\Phi_k\rangle\rightarrow |\psi_2\rangle^{\otimes m}\otimes
|\Phi_k\rangle)\geq p_1p_2p_3.
\end{equation}

As we just mentioned above,  a maximally entangled state cannot be
used to catalyze any transformation, which can be treated as  a
direct consequence of Vidal's formula. Thus,
\begin{equation}\label{bell}
\begin{array}{l}
p_{max}(|\psi_1\rangle^{\otimes m}\otimes
|\Phi_k\rangle\rightarrow |\psi_2\rangle^{\otimes m}\otimes
|\Phi_k\rangle)\\
\\
=p_{max}(|\psi_1\rangle^{\otimes m}\rightarrow
|\psi_2\rangle^{\otimes m}) {\rm\ for\ any\ } m\geq 1.
\end{array}
\end{equation}

Combining Eqs. (\ref{lowbound}) and (\ref{bell}), we finally have
\begin{equation}\label{pmp1p2p3}
p_{max}(|\psi_1\rangle^{\otimes m}\rightarrow
|\psi_2\rangle^{\otimes m})\geq p_1p_2p_3,
\end{equation}
Substituting Eqs. (\ref{p1}), (\ref{p2}) and (\ref{p3}) into Eq.
(\ref{pmp1p2p3}) and taking average yield that
\begin{equation}\label{elocclessmlocc}
\begin{array}{l}
[p_{max}(|\psi_1\rangle^{\otimes m}\rightarrow
|\psi_2\rangle^{\otimes m})]^{\frac{1}{m}}\\
\\
\geq (k\gamma_k)^{\frac{1}{m}} p_{max}(|\psi_1\rangle\otimes
|\phi\rangle\rightarrow|\psi_2\rangle\otimes |\phi\rangle).
\end{array}
\end{equation}
The above equation has an interesting physical meaning: the
success probability of  simulating a catalyst $|\phi\rangle$ by an
$m$-copy transformation has a lower bound
$(k\gamma_k)^{\frac{1}{m}}$. Taking supremum according to $m$
yields
\begin{equation}\label{mcatalyst}
P_M\geq p_{max}(|\psi_1\rangle\otimes |\phi\rangle\rightarrow
|\psi_2\rangle\otimes |\phi\rangle).
\end{equation}
Here we have used a well known result in elementary calculus:
\begin{equation}\label{limit}
\sup\limits_m a^{\frac{1}{m}}=1, {\rm\ for\ any\ }0<a\leq 1,
\end{equation}
where the supremum $1$ can be attained asymptotically when $m$
tends to infinity. This fact will be useful latter.

We thus finish the simulation of the  catalyst state
$|\phi\rangle$ by means of MLOCC asymptotically. Now the desired
result $P_E\leq P_M$ follows directly by taking supremum
according to $|\phi\rangle$ in the right-hand side of Eq. (\ref{mcatalyst}). \hfill $\square$\\

As we mentioned above, a direct application of Theorem \ref{mise}
is to evaluate the optimal conversion probability of an ELOCC
transformation. To be specific, let $|\psi_1\rangle$ and
$|\psi_2\rangle$ be two given $n\times n$ states. First, we show
that the geometric average conversion probability of $m$-copy
transformation $|\psi_1\rangle^{\otimes m}\rightarrow
|\psi_2\rangle^{\otimes m}$, or shortly, $P_M^{(m)}$, can be
computed  within the polynomial time of $m$ when $n$ is fixed. It
is easy to check that both $|\psi_1\rangle^{\otimes m}$ and
$|\psi_2\rangle^{\otimes m}$ have at most $\choose{n-1+m}{n-1}$
distinct Schmidt coefficients. Then one only needs
$O(\choose{n-1+m}{n-1}\log \choose{n-1+m}{n-1})$ time to sort the
Schmidt coefficients of $|\psi_1\rangle^{\otimes m}$ and
$|\psi_2\rangle^{\otimes m}$ into nonincreasing order,
respectively. By Vidal's formula, the geometric average conversion
probability of $m$-copy transformation can be calculated
efficiently. A more careful analysis shows that the time
complexity is about $O(\choose{n-1+m}{n-1}\log
\choose{n-1+m}{n-1})$, which is essentially the same as
$O(m^{n-1}\log m)$ when $n$ is fixed. Second, by Eq. (\ref{pm}),
we can obtain an approximation of $P_M$ as accurate as possible by
calculating $P_M^{(m)}$ for a large $m$ since $P_M^{(m)}$
converges to $P_M$ when $m$ tends to infinity (this can be seen
from Eq. (\ref{limit}) and the line after it).  Third, by Theorem
\ref{mise}, such an approximation of $P_M$ can also be used to
approximate $P_E$.

In the proof of Theorem \ref{mise}, we notice that any
multiple-copy transformation can be simulated by an
entanglement-assisted transformation in a finite manner. However,
the proof of the converse part, i.e., simulating ELOCC by MLOCC,
is only given in an asymptotical way. Thereby a natural question
is to ask whether one can design a stronger protocol which can
simulate an entanglement-assisted transformation with a finite
dimensional catalyst by using a multiple-copy entanglement
transformation in a finite manner. The second example proposed
above indicates that such a finite simulation is not possible for
certain catalyst state. Interestingly, except for a special case,
any finite dimensional catalyst can be simulated by a
multiple-copy transformation in a finite manner.
\begin{theorem}\label{probme}\upshape
For any transformation $|\psi_1\rangle\rightarrow |\psi_2\rangle$
and a catalyst $|\phi\rangle$, denote
$p=p_{max}(|\psi_1\rangle\otimes |\phi\rangle\rightarrow
|\psi_2\rangle\otimes |\phi\rangle)$. If $p<{\rm min}\{1,
\frac{\alpha_n}{\beta_n}\}$,  then there exists a finite positive
integer $m$ such that $p_M^{(m)}(|\psi_1\rangle\rightarrow
|\psi_2\rangle)\geq p$, where $\alpha_n$ and $\beta_n$ denote  the
least Schmidt coefficients of $|\psi_1\rangle$ and
$|\psi_2\rangle$, respectively.
\end{theorem}
In other words, the geometric average value of the probability of
(single-copy) transformation $|\psi_1\rangle\rightarrow
|\psi_2\rangle$ in the environment of $m$-copy transformation
$|\psi_1\rangle^{\otimes m}\rightarrow |\psi_2\rangle^{\otimes m}$
is not less than $p$. This means that under the assumption,  the
role of the catalyst $|\phi\rangle$ can be replaced by an $m$-copy
transformation.

\textrm{Proof.} First, notice that Theorem 5 in Ref. \cite{FDY04}
provides a necessary and sufficient condition of when ELOCC is
more powerful than mere LOCC. Applying this result to the
transformation $|\psi_1\rangle\otimes |\phi\rangle\rightarrow
|\psi_2\rangle\otimes |\phi\rangle$ yields
\begin{equation}\label{eq1}
P_E(|\psi_1\rangle\otimes
|\phi\rangle\rightarrow|\psi_2\rangle\otimes |\phi\rangle)>p {\rm\
\ iff\ \ }
 p<{\rm min}\{1,
\frac{\alpha_n\gamma_k}{\beta_n\gamma_k}\},
\end{equation}
where  $\gamma_k$ is the least Schmidt coefficient of
$|\phi\rangle$.

Second, by the definition of
$P_E(|\psi_1\rangle\rightarrow|\psi_2\rangle)$, we have
\begin{equation}\label{pecatalyst}
P_E(|\psi_1\rangle\rightarrow|\psi_2\rangle)=P_E(|\psi_1\rangle\otimes
|\phi\rangle\rightarrow|\psi_2\rangle\otimes |\phi\rangle).
\end{equation}

Third, by Theorem \ref{mise}, we have
$P_E(|\psi_1\rangle\rightarrow|\psi_2\rangle)=P_M(|\psi_1\rangle\rightarrow|\psi_2\rangle).$
This together with Eqs. (\ref{eq1}) and (\ref{pecatalyst})  yields
that
\begin{equation}\label{eq5}
P_M(|\psi_1\rangle\rightarrow|\psi_2\rangle)>p {\rm\ \ iff\ \ }
p<{\rm min}\{1,\frac{\alpha_n}{\beta_n}\}.
\end{equation}
The condition on right-hand side of Eq. (\ref{eq5}) is fulfilled
by the assumption. Thus we have
$P_M(|\psi_1\rangle\rightarrow|\psi_2\rangle)>p$. By the
definition of $P_M(|\psi_1\rangle\rightarrow|\psi_2\rangle)$,
there exists a finite positive integer $m$ such that
$$[p_{max}(|\psi_1\rangle^{\otimes m}\rightarrow
|\psi_2\rangle^{\otimes m})]^{\frac{1}{m}}\geq p.$$ This completes
the proof of the theorem. \hfill $\square$\\

It seems to be a challenging problem to characterize when  a
catalyst $|\phi\rangle$ can be simulated by a finite-copy
transformation in the case of $p={\rm
min}\{1,\frac{\alpha_n}{\beta_n}\}$.

In conclusion, we have examined the relationship between
entanglement-assisted transformation and multiple-copy
transformation, and have proved that these two ways of
manipulation of bipartite pure states are equivalent in the sense
they can simulate each other's ability to implement a desired
transformation from a given source state to a given target state
with the same optimal conversion probability. It would be
interesting to obtain  a similar equivalence between
entanglement-assisted transformation and multiple-copy
entanglement transformation  for pure states shared by three or
more parties, and also to extend our results presented here to
mixed states.

The authors wish to acknowledge the colleagues in the Quantum
Computation and Quantum Information Research Group for many useful
discussions. This work was partly supported by the Natural Science
Foundation of China (Grant Nos: 60273003, 60433050, 60321002, and
60305005). R. Duan acknowledges the financial support of Tsinghua
University (Grant No. 052420003).

\end{document}